\documentclass{aastex631}
\usepackage{graphicx}
\usepackage[normalem]{ulem}
\usepackage{float}
\usepackage[caption=false]{subfig}
\usepackage{enumitem}
\useunder{\uline}{\ul}{}
\usepackage{tabularray}
\usepackage{csquotes}

\begin{document}

\title{NIRCam Performance on JWST In Flight}


\author [0000-0002-7893-6170] {Marcia J. Rieke}
\affiliation{Steward Observatory, University of Arizona, 933 N. Cherry Ave,
Tucson, AZ 85721, USA}

\correspondingauthor{Marcia J. Rieke}
\email{mrieke@arizona.edu}



\author{Douglas M. Kelly}
\affiliation{Steward Observatory, University of Arizona, 933 N. Cherry Ave,
Tucson, AZ 85721, USA}

\author{Karl Misselt}
\affiliation{Steward Observatory, University of Arizona, 933 N. Cherry Ave,
Tucson, AZ 85721, USA}

\author{John Stansberry}
\affiliation{Space Telescope Science Institute, 3700 San Martin Drive,
Baltimore, MD 21218}

\author{Martha Boyer}
\affiliation{Space Telescope Science Institute, 3700 San Martin Drive,
Baltimore, MD 21218}

\author{Thomas Beatty}
\affiliation{Department of Astronomy,
University of Wisconsin,
Madison, Madison, WI 53706}

\author{Eiichi Egami}
\affiliation{Steward Observatory, University of Arizona, 933 N. Cherry Ave,
Tucson, AZ 85721, USA}

\author{Michael Florian}
\affiliation{Steward Observatory, University of Arizona, 933 N. Cherry Ave,
Tucson, AZ 85721, USA}

\author[0000-0002-8963-8056]{Thomas P. Greene}
\affiliation{NASA Ames Research Center, Space Science and Astrobiology Division,
MS 245-6, Moffett Field, CA, 94035, USA}

\author[0000-0003-4565-8239]{Kevin Hainline}
\affiliation{Steward Observatory, University of Arizona, 933 N. Cherry Ave,
Tucson, AZ 85721, USA}

\author[0000-0002-0834-6140]{Jarron Leisenring}
\affiliation{Steward Observatory, University of Arizona, 933 N. Cherry Ave,
Tucson, AZ 85721, USA}

\author{Thomas Roellig}
\affiliation{NASA Ames Research Center, Space Science and Astrobiology Division,
MS 245-6, Moffett Field, CA, 94035, USA}

\author[0000-0001-8291-6490]{Everett Schlawin}
\affiliation{Steward Observatory, University of Arizona, 933 N. Cherry Ave,
Tucson, AZ 85721, USA}

\author[0000-0002-4622-6617]{Fengwu Sun}
\affiliation{Steward Observatory, University of Arizona, 933 N. Cherry Ave,
Tucson, AZ 85721, USA}

\author{Lee Tinnin}
\affiliation{Steward Observatory, University of Arizona, 933 N. Cherry Ave,
Tucson, AZ 85721, USA}

\author{Christina C. Williams}
\affiliation{National Optical-Infrared Research Laboratory , 950 N Cherry Ave,
Tucson, AZ 85719}

\author[0000-0001-9262-9997]{Christopher N. A. Willmer}
\affiliation{Steward Observatory, University of Arizona, 933 N. Cherry Ave,
Tucson, AZ 85721, USA}

\author{Debra Wilson}
\affiliation{Steward Observatory, University of Arizona, 933 N. Cherry Ave,
Tucson, AZ 85721, USA}

\author {Charles R. Clark}
\affiliation{ Goddard Space Flight Center,
Greenbelt, MD }

\author {Scott Rohrbach}\
\affiliation{ Goddard Space Flight Center,
Greenbelt, MD }

\author{Brian Brooks}
\affiliation{Space Telescope Science Institute, 3700 San Martin Drive,
Baltimore, MD 21218}

\author{Alicia Canipe}
\affiliation{Space Telescope Science Institute, 3700 San Martin Drive,
Baltimore, MD 21218}

\author{Matteo Correnti}
\affiliation{INAF, Osservatorio Astronomico di Roma, via Frascati 33, 00078, Monteporzio Catone, Rome, Italy}
\affiliation{ASI-Space Science Data Center, Via del Politecnico, I-00133, Rome, Italy}

\author{Audrey DiFelice}
\affiliation{Space Telescope Science Institute, 3700 San Martin Drive,
Baltimore, MD 21218}

\author{Mario Gennaro}
\affiliation{Space Telescope Science Institute, 3700 San Martin Drive,
Baltimore, MD 21218}

\author{Julian Girard}
\affiliation{Space Telescope Science Institute, 3700 San Martin Drive,
Baltimore, MD 21218}

\author{George Hartig}
\affiliation{Space Telescope Science Institute, 3700 San Martin Drive,
Baltimore, MD 21218}

\author{Bryan Hilbert}
\affiliation{Space Telescope Science Institute, 3700 San Martin Drive,
Baltimore, MD 21218}

\author[0000-0002-6610-2048]{Anton M. Koekemoer}
\affiliation{Space Telescope Science Institute, 3700 San Martin Drive,
Baltimore, MD 21218}

\author{Nikolay K. Nikolov}
\affiliation{Space Telescope Science Institute, 3700 San Martin Drive,
Baltimore, MD 21218}

\author{Norbert Pirzkal}
\affiliation{Space Telescope Science Institute, 3700 San Martin Drive,
Baltimore, MD 21218}

\author[0000-0002-4410-5387]{Armin Rest}
\affiliation{Space Telescope Science Institute, 3700 San Martin Drive,
Baltimore, MD 21218}
\affiliation{Department of Physics and Astronomy, Johns Hopkins University, 
Baltimore, MD 21218 }

\author{Massimo Robberto}
\affiliation{Space Telescope Science Institute, 3700 San Martin Drive,
Baltimore, MD 21218}
\affiliation{Johns Hopkins University, 3400 N. Charles Street, Baltimore, MD 21218, USA}

\author{Ben Sunnquist}
\affiliation{Space Telescope Science Institute, 3700 San Martin Drive,
Baltimore, MD 21218}

\author{Randal Telfer}
\affiliation{Space Telescope Science Institute, 3700 San Martin Drive,
Baltimore, MD 21218}

\author{Chi Rai Wu}
\affiliation{Space Telescope Science Institute, 3700 San Martin Drive,
Baltimore, MD 21218}

\author{Malcolm Ferry}
\affiliation{ Lockheed Martin Advanced Technology Center, 
3251 Hanover St.,
Palo Alto, CA 94304}

\author{Dan Lewis}
\affiliation{ Lockheed Martin Advanced Technology Center, 3251 Hanover St.,
Palo Alto, CA 94304}

\author{Stefi Baum}
\affiliation{Faculty of Science, 230 Machray Hall, 186 Dysart Road,
University of Manitoba,
Winnipeg, MB Canada R3T 2N2}

\author[0000-0002-5627-5471]{Charles Beichman}
\affiliation{NASA Exoplanet Science Institute/IPAC,
Jet Propulsion Laboratory, California  Institute of Technology,
1200 E California Blvd,
Pasadena, CA 91125}

\author{René Doyon}
\affiliation{Département de physique, Université de Montréal,
C.P. 6128, Succursale Centre-Ville,
Montréal, QC Canada   H3C 3J7}

\author{Alan Dressler}
\affiliation{The Observatories, The Carnegie Institution for Science,
813 Santa Barbara St.,
Pasadena, CA 91101, USA}

\author[0000-0002-2929-3121]{Daniel J.\ Eisenstein}
\affiliation{Center for Astrophysics $|$ Harvard \& Smithsonian, 60 Garden St., Cambridge MA 02138 USA}

\author{Laura Ferrarese}
\affiliation{National Research Council Canada, Herzberg Astronomy and Astrophysics,
5071 West Saanich Rd.,
Victoria, BC, Canada V9E 2E7 }

\author[0000-0003-0786-2140]{Klaus Hodapp}
\affiliation{Institute for Astronomy, 
640 N Aohoku Pl
Hilo, HI 96720, USA }

\author{Scott Horner}
\affiliation{881 Spinosa Dr., Sunnyvale, CA 94087}

\author{Daniel T. Jaffe}
\affiliation{The University of Texas at Austin,
Department of Astronomy RLM 16.342,
Austin, TX 78712}

\author{Doug Johnstone}
\affiliation{National Research Council Canada,
Herzberg Astronomy and Astrophysics,
5071 West Saanich Rd.,
Victoria, BC, Canada V9E 2E7 }
\affiliation{Department of Physics and Astronomy, University of Victoria, Victoria, BC, V8P 5C2, Canada}

\author{John Krist}
\affiliation{Jet Propulsion Laboratory,
4800 Oak Grove Drive M/S 183-900,
Pasadena, CA 91109 }

\author{Peter Martin}
\affiliation{Canadian Institute for Theoretical Astrophysics,
University of Toronto, 
McLennan Physical Laboratories,
60 St. George Street, 
Toronto, Ontario, Canada M5S 3H8 }

\author{Donald W. McCarthy}
\affiliation{Steward Observatory, University of Arizona, 933 N. Cherry Ave,
Tucson, AZ 85721, USA}

\author{Michael Meyer}
\affiliation{Department of Astronomy,
University of Michigan,
1085 S. University,
Ann Arbor, MI 48109 }

\author[0000-0003-2303-6519]{George H. Rieke}
\affiliation{Steward Observatory, University of Arizona, 933 N. Cherry Ave,
Tucson, AZ 85721, USA}

\author{John Trauger}
\affiliation{Jet Propulsion Laboratory,
4800 Oak Grove Drive M/S 183-900,
Pasadena, CA 91109 }

\author{Erick T. Young}
\affiliation{ USRA,
Mountain View, CA, USA }

\begin{abstract}

The Near Infrared Camera for the James Webb Space Telescope is delivering the imagery that astronomers have hoped for ever since JWST was proposed back in the 1990s. In the Commissioning Period that extended from right after launch to early July 2022 NIRCam has been subjected to a number of performance tests and operational checks. The camera is exceeding pre-launch expectations in virtually all areas with very few surprises discovered in flight. NIRCam also delivered the imagery needed by the Wavefront Sensing Team for use in aligning the telescope mirror segments (\citealt{Acton_etal2022}, \citealt{McElwain_etal2022}). 

\end{abstract}


\keywords{infrared camera, HgCdTe detector}

\section{Introduction} \label{sec:intro}
All of the original concepts for what we now call the James Webb Space Telescope (JWST)
included a relatively large field-of-view camera spanning at least the $1$ to $5 \mu m$ range. NIRCam's science role was initially focused on deep surveys to find the most distant galaxies (\citealt{Stockman_1997}), and of course, an instrument capable
of finding these galaxies is also capable of many other observations. The deep survey use drove the need for as large a
field of view as possible since no one knew how common galaxies at $z>10$ would be at the time NIRCam was being designed. 
What was known is that photometric redshifts would be a powerful tool and so NIRCam would need a filter set that spanned
its entire wavelength range, with filter widths that needed to be optimized between sensitivity and wavelength
discrimination. 

Over time the 
role of NIRCam expanded to include providing the imaging used for wavefront sensing and telescope alignment, which 
rendered NIRCam a mission-critical element. The final NIRCam design incorporates two copies (modules) of the camera to provide 
full block redundancy to maximize availability of the wavefront sensing capability. Use of NIRCam as the wavefront sensor 
also meant that NIRCam must have very small intrinsic wavefront error so it does not imprint errors on the other 
instruments (\cite{Acton_etal2004}, \cite{Acton_etal2018}). 

\begin{figure}[ht!]
\plotone{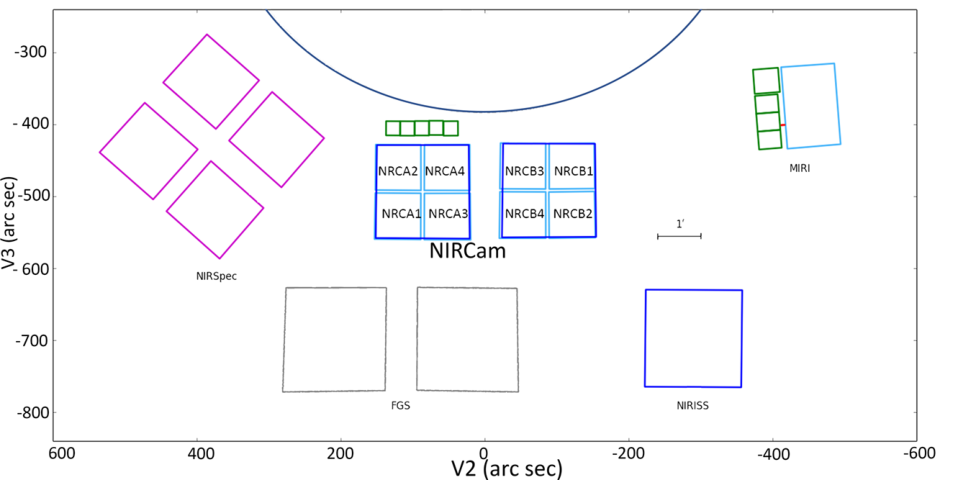}
\caption{NIRCam's location in the JWST field of view. Designations for the short wavelength arrays are shown with the long wavelength arrays called NRCALONG and NRCBLONG with the fields of view outlined in blue. The long wavelength arrays cover the same area as the four short wavelength arrays in a module. The small green squares show the locations of the coronagraphic fields. V2 and V3 refer to the telescope coordinate system
where V3 points away from the sunshield and V2 is perpendicular to V3 to form a right-handed coordinate system.}
\label{fig:fov}
\end{figure}

In the final NIRCam design, two modules provide 2.2 arc min x 2.2 arc min fields of view each as shown in 
Figure \ref{fig:fov}. Each field of view can be observed using two filters simultaneously through the use of dichroics that split the wavelength range into $0.6\mu m$ to $2.3\mu m$ and $2.4\mu m$ to $5\mu m$ with a deadband from $2.3\mu m$ to $2.4\mu m$, effectively doubling
the field for programs needing full filter coverage. NIRCam also includes 
components to enable coronagraphy; these do not affect the area available for direct imaging. Grisms in the long wavelength arms, originally included to expand the coarse phasing capture range, can be used to provide a time-series spectroscopic capability useful for exoplanet observations and to provide wide field spectroscopy, a powerful took for finding emission-line galaxies.

Lockheed Martin's Advanced Technology Center led the design and construction of NIRCam. The University of Arizona Team
developed the focal plane assemblies in collaboration with Teledyne Imaging Systems. Arizona also developed the
initial pipeline and reference files which were used by Space Telescope Science Institute in producing the first 
flight pipeline. The Space Telescope Science Institute maintains documentation about the telescope and instruments including 
NIRCam at https://jwst-docs.stsci.edu/.

\section{Design and Construction} 
The NIRCam design was driven by wanting a large field of view for efficient surveys and by the wavefront error requirement imposed by its use as the mission wavefront sensor. NIRCam is required to have less than 90 nm of absolute wavefront error at $2.12 \mu m$. The ground-to-flight change in wavefront error is required to be less than 40 nm (\citealt{Huff_2005}). The redundancy imposed by being the wavefront sensor meshes well with the efficient
survey need as having two complete copies of NIRCam doubles the field of view. Assembly and test of the two 
modules was facilitated by including a tip-tilt focus mechanism in each module. This mechanism can be used to
align the NIRCam pupil with the telescope as required for wavefront sensing and the coarse phasing step of
aligning the telescope. The ability to adjust the focus of each module separately relaxed the need for
each module to have precisely the same focus. The fields of view of the two modules are separated by $\sim 44$ arc sec and cover a total of 9.7 square arc minutes for one wavelength.

\begin{figure}[ht!]
\plotone{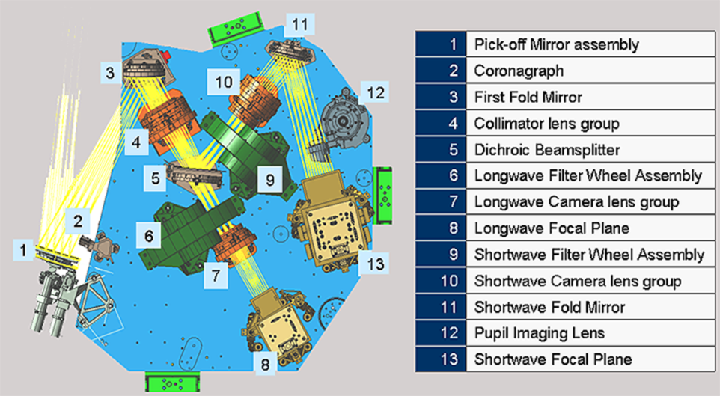}
\caption{One half of NIRCam showing its optical train and mechanisms. Starlight from the telescope enters the instrument from the upper left, enters the module after reflection from the pick-off mirror (1) and then reflects into the main body of the camera where the dichroic beamsplitter (5) separates the long and short wavelength light. The other module is a mirror image of the one shown here, mounted back-to-back behind what is shown.}
\label{fig:instrument}
\end{figure}

The survey use drove the choice 
of using dichroics to split the light into short and long wavelength optical trains in each of the redundant
NIRCam modules. Figure \ref{fig:instrument} presents the light path and layout of one module. Note that this figure shows the pupil imaging lens in the deployed position rather than the usual retracted position. The wavelength split also facilitates the matching of pixel scales to diffraction-limited sampling so 
NIRCam provides Nyquist sampling of the point spread function at 2 $\mu m$ and at 4 $\mu m$. The overlap between the short wavelength and long wavelength fields of view is $\sim 96\%$ in both modules. Each arm in
a module has two wheel mechanisms mounted together with one wheel carrying mainly filters and the other wheel
carrying optical elements that need to be used at a pupil as well some additional filters. The wheels are referred to as the filter wheel and the pupil wheel. They are moved independently.

Another benefit 
of the short wavelength / long wavelength split is that detector arrays (referred to as SCAs for Sensor Chip Assemblies) with different cut-off wavelengths 
and different anti-reflection coatings could be used. Teledyne Imaging Systems fabricated the arrays for NIRCam.  
NIRCam uses eight 2.5 $\mu m$ cut-off SCAs and two 5 $\mu m$ cut-off SCAs. It is somewhat easier to make the 
2.5-$\mu m$ devices than the 5-$\mu m$ devices so the wavelength split eased the production process for this relatively 
large number of arrays.  Last, the wavelength split also eased the blocking requirements on NIRCam's bandpass 
filters. The short wavelength filters do not need blocking beyond the response band of the 2.5-$\mu m$ detectors. The
dichroics have silicon substrates so the long wavelength filters do not need blocking below $1.08 \mu m$. 

A key design choice for NIRCam was the decision to use a refractive design. This choice was motivated by such a 
design yielding a more compact and less massive instrument. Some of the very early designs for NIRCam using reflective
optics yielded an instrument so large that there may not have been room and mass for a fourth instrument on JWST. The refractive
design choice also helped enable the fully redundant design. 

\section{Wavefront Sensing} \label{sec:wfs}
\begin{figure}[htbp!]
\plotone{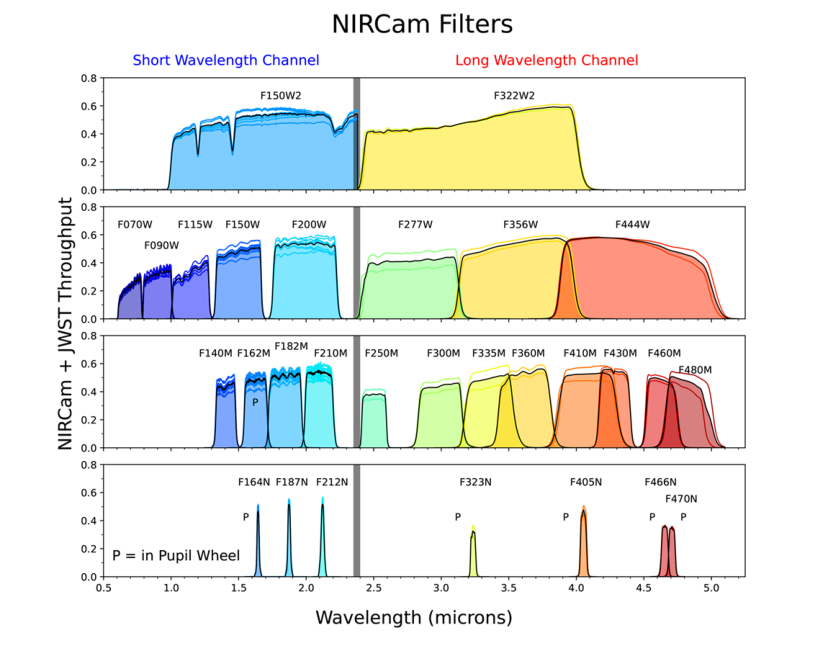}
\caption{Throughputs of NIRCam's filters. All terms affecting throughpout 
including detector quantum efficiency and telescope reflectivity are included. The gray bar denotes the dichroic deadband. The letter P indicates filters that are mounted in the pupil wheels.}
\label{fig:filters}
\end{figure}

The JWST primary mirror has eighteen fully adjustable hexagonal segments that must be aligned to perform 
like a single, monolithic mirror. NIRCam was used to locate all of the segment images after launch and provided
the imagery for the initial mirror alignment. The first step in phasing the eighteen segments after segment-level adjustments were completed using dispersed Hartmann sensors (DHSes) mounted in the short wavelength pupil wheel to provide spectra at $R\sim300$
resulting from the interference of two mirror segments. The F150W2 filter includes two notches in its spectral
response shown in Figure \ref{fig:filters} which provides the wavelength calibration needed to convert the interference pattern into a piston difference between the segments. Each DHS spans ten segment pairs and produces ten spectra. The two DHSes are mounted at 60 degrees with respect to each other to provide 20 separate measurements to adjust the segments in piston. Relatively large focus steps between segments can be 
measured using these data, and were a key step in the initial mirror alignment but not needed for routine wavefront
sensing. 

Every two days NIRCam takes images for focus diverse phase retrieval to monitor the alignment of the primary mirror. Mirror adjustments are then executed when the alignment has drifted to where the overall observatory wavefront error 
has increased to  $\sim80$ nm. These adjustments have been needed on average about every 17 days. Focus diversity is achieved
by using weak lenses in the short wavelength arm with one mounted in the filter wheel and two in the pupil wheel. These lenses defocus the star light by either +/- 4 waves, +/- 8 waves, and as much as +12 waves using +4 and +8 in series. The wavefront sensing is executed at a wavelength of $2.1 \mu m$.  Two of these weak lenses, +4 and +8, can
also be used to defocus stars observed in time series measurements (see \ref{subsec:tso}).
Focus diverse phase retrieval and the overall wavefront sensing process are described in detail in \cite{Acton_etal2012}, \cite{Perrin_etal2016}, and \cite{Acton_etal2022}.

\section{Observing Modes} \label{sec:modes}
NIRCam has five observing modes. Two of these modes, grism time series and photometric time series, are
very similar. Table \ref{table:modes} presents the basic characteristics of the modes. The imaging mode
serves as the basis for all five modes with its collection of data from both short wavelength 
and long wavelength arms simultaneously and with detectors read out using sampling up the ramp.
The wide field slitless spectroscopic (WFSS) mode includes acquisition of direct images in addition
to the grism images to enable source detection and wavelength calibration in the spectroscopic images.
In this mode the short wavelength arm is used for direct imaging. For grism time series, the long wavelength
arm uses a grism in series with a bandpass filter for out of band blocking. The short wavelength arm is used with a choice of weak lens and the appropriate $\sim2 \mu m$ filter. The photometric time series mode uses bandpass filters in Module B. In coronagraphic mode, a target star is placed behind an occulter and images are acquired.

\begin{table}
\center{}
\caption{NIRCam Observing Modes}
\begin{tblr}{|c|c|c|c|}
\hline
Mode & {Number of \\ SCAs Used} & Typical Use & Special Requirements \\
\hline 
Imaging                 & 2,5,or 10           & General 0.7 to 5  $\mu m$  imaging                 & {Subarrays available, RAPID \\ mode using 10 SCAs only for \\single exposures} \\ 
\hline
 WFSS                    & {10 \\ (2 for spectra, 8 for imaging)}                & {Wide field emission line survey \\ 2.4 to 5  $\mu m$ with\\ short wavelength imaging} & {Careful planning needed for \\ full spatial coverage }  \\ 
 \hline
 Grism Time Series       & 2 (Mod A only)      & Transit spectroscopy 2.4 to 5 $\mu m$    & {Stripe mode can extend \\bright limit, target \\ acquisition may be desirable}   \\
 \hline
 {Photometric Time \\ Series} & 2 or 5 (Mod B only) & High time resolution photometry  & \\  
 \hline
 Coronagraphy            & 2 (Mod A only)      & High contrast imaging  & Target acquisition required \\
 \hline
\end{tblr}
\label{table:modes}
\end{table}

\subsection{Imaging Mode} \label{subsec:phot_mode}
As described earlier, NIRCam's basic imaging mode is used for both science imaging and for wavefront sensing 
imaging. Both short wavelength and long wavelength arms are read out simultaneously, 
but if a point source or limited area is to  be observed, only one module of NIRCam 
can be used to save on data volume. For surveys, both modules can be used. There is no hardware
requirement that the two modules be configured with the same filter choices although the Astronomer's 
Proposal Tool does not support different filters currently. The modules are 
constrained to use the same readout patterns. If a bright source is observed, subarrays 
can be used to extend the camera's dynamic range but all arrays must be read out using 
the same size subarray so typically only one short wavelength array and the corresponding 
long wavelength array will be read out. In the case of extended bright sources, one long wavelength and a set of four short wavelength subarrays at the center of the B module, with approximately overlapping areas are used. For point sources, one short wavelength and one long wavelength subarray can be configured, located at the upper left corner of the NRCB1, a short wavelength SCA. Figure \ref{fig:fov} shows the location of NIRCam and its SCAs in the JWST focal plane.

The throughput of the telescope plus NIRCam system is higher than pre-launch predictions. 
This improvement is partly the result of arriving at L2 with fewer particulates and other contaminants such as ice and partly the result of conservatism in the pre-launch estimates. 
Figure \ref{fig:filters} shows the throughputs which are remarkable for a camera employing six lenses, a tribute to the quality of the lens anti-reflection coatings. The throughput has been measured by comparing the known flux of calibration stars
to the detected flux in a 2.5 arc second radius aperture. The detected flux was corrected to the total flux
using an encircled energy curve based on a PSF computed using the optical path difference map for the date 
of the observation available from the data archive MAST. No correction was made for gaps or strut obscuration in the telescope throughput.

\begin{figure}[ht!]
  \centering
  \includegraphics[width=9cm]{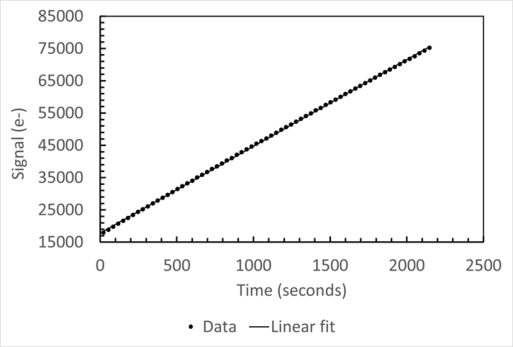}\hfill 
  \includegraphics[width=9cm]{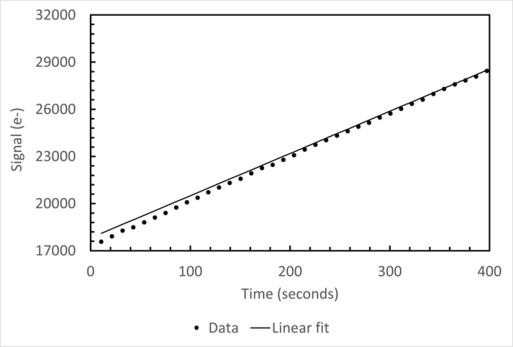}
  \caption{On the left, a ramp for a pixel illuminated by a star showing the linear behavior of the charge collection. On the right, the low signal portion of the ramp illustrates the non-linearity present at low signal levels. }
  \label{fig:ramps}
\end{figure}

\begin{figure}[ht!]
\centering
\includegraphics[width=9cm]{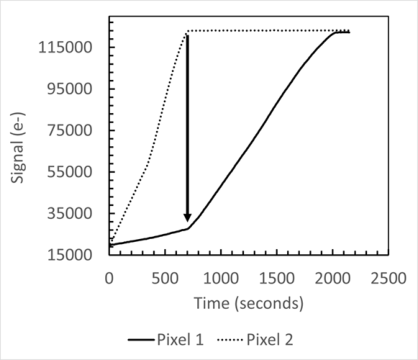}
\caption{Ramps from adjacent pixels illustrating how the slope changes when the 
adjacent pixel saturates and charge migration begins.}
\label{fig:saturated}
\end{figure}

Figure \ref{fig:ramps} displays a pixel ramp from data taken during a Cycle 1 calibration program, illustrating how charge builds up over time from a source and is very linear. The start of the ramp lies below the linear fit, which is indicative of charge being lost to traps in the detector material. This 
loss of charge to traps could affect low signal observations but in many cases there is sufficient
charge collection from the zodiacal background that this is not an issue. Figure \ref{fig:saturated} displays a ramp that reaches saturation, and also shows how a 
pixel that reaches saturation affects the slope of an adjacent pixel as charge migrates from the saturated pixel
to adjacent pixels. The migration of saturated charge is mainly a bright object problem but can complicate
extraction of faint sources on the fringes of a bright PSF. The pipeline includes a saturation check and a linearity correction step to compensate for the change in gain of the readout circuit as the charge on the integrating node increases (see \citealt{Plazas_etal2017}).  

NIRCam is background-limited through wide filters as predicted before launch. Table \ref{table:sensitivity} gives the current sensitivity for long exposures. The performance of NIRCam's detectors varies from SCA to SCA somewhat. Here we will use instrument-wide averages unless denoted otherwise. The flight read noise and dark current of the NIRCam short wavelength SCAs are slightly higher than measured in ground 
testing which is undoubtedly the result of cosmic-ray hits. The eight short wavelength SCAs had an average read noise of 6.2 electrons and the two long wavelength SCAs had an average of 9.1 electrons in 1000 seconds in ground testing. In flight, these two values changed to 6.4 and 7.5 electrons respectively. The long wavelength SCAs benefit from being surrounded by a colder environment in flight so that the effect of small light leaks in the NIRCam baffles is mitigated. 

\begin{table}
\center{}
\caption{NIRCam sensitivity 10 $\sigma$ in 10,000 second in 2.5-pixel radius aperture with a background 1.2x the minimum zodaical light level}
\begin{tblr}{|c|c|c|c|c|c|c|}
\hline
Filter & {$\lambda_{pivot}$ \\ $\mu m$ }  & nJy &  & Filter & {$\lambda_{pivot}$ \\ $\mu m$ } & nJy\\
\hline 
 F070W & 0.704 & 11.85 &  & F277W & 2.786 & 10.59 \\
\hline
 F090W & 0.901 &  9.74 &  & F356W & 3.563 &  8.88 \\
 \hline
 F115W & 1.154 &  8.57 &  & F444W & 4.421 & 17.31 \\
 \hline
 F150W & 1.501 &  6.99 &  & F250M & 2.503 & 24.63 \\
 \hline
 F200W & 1.990 &  6.22 &  & F300M & 2.996 & 16.75  \\
 \hline
 F140M & 1.404 & 11.85 &  & F335M & 3.365 & 14.91  \\
\hline
 F162M & 1.626 & 11.35 &  & F360M & 3.621 & 15.14  \\
 \hline
 F182M & 1.845 &  9.28 &  & F410M & 4.092 & 18.74 \\
 \hline
 F210M & 2.093 & 11.56 &  & F430M & 4.280 & 35.14 \\
 \hline
 F164N & 1.644 & 69.08 &  & F460M & 4.624 & 56.81 \\
 \hline
 F187N & 1.874 & 65.00 &  & F480M & 4.834 & 63.75 \\
 \hline
 F212N & 2.120 & 70.31 &  & F323N & 3.237 & 113.73 \\
 \hline
 F405N & 4.055 & 103.44 &  & F466N & 4.654 & 173.14 \\
 \hline
 F470N & 4.707 & 199.80 &  &  &  &\\
 \hline
\end{tblr}
\label{table:sensitivity}
\end{table}

\begin{figure}[ht!]
  \centering
  \includegraphics[width=9cm]{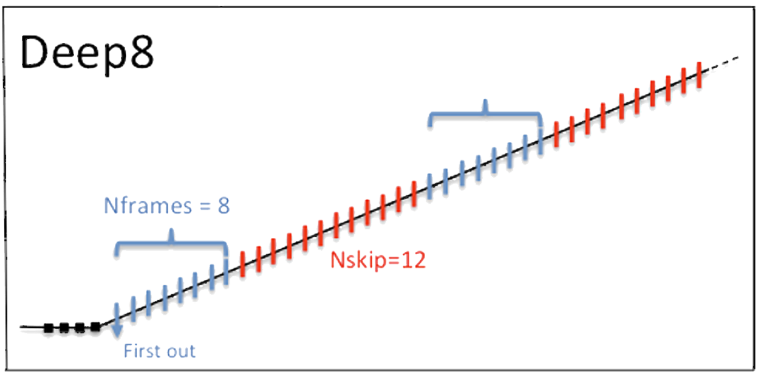}
  \caption{The DEEP8 readout pattern illustrating the co-addition of eight samples and skipping of reads to minimize data volume. }
  \label{fig:deep8}
\end{figure}

Because NIRCam generates
a large volume of data when all ten SCAs are read out, a variant of sampling up the ramp is used where two, 
four, or eight samples are co-added and then a number of reads are clocked but not included in the
co-adds. Figure \ref{fig:deep8} illustrates the DEEP8 readout pattern which can be used for long exposures with all ten 
SCAs and still have data volume available for data acquisition in parallel such as imaging with MIRI or NIRISS or spectra with NIRSpec. Other readout patterns with more samples up the ramp can be used when data volume 
is not a limitation.

Any of the readout patterns with co-addition also include saving of the first read separately to recover any
pixels that reach saturation in subsequent reads. Currently the Space Telescope Science Institute pipeline is not using this \enquote{Zero read frame} but will do so in the future.

\subsection{Wide Field Slitless Spectroscopy (WFSS)} 
\label{subsec:wfss}
\begin{figure}[ht!]
\plotone{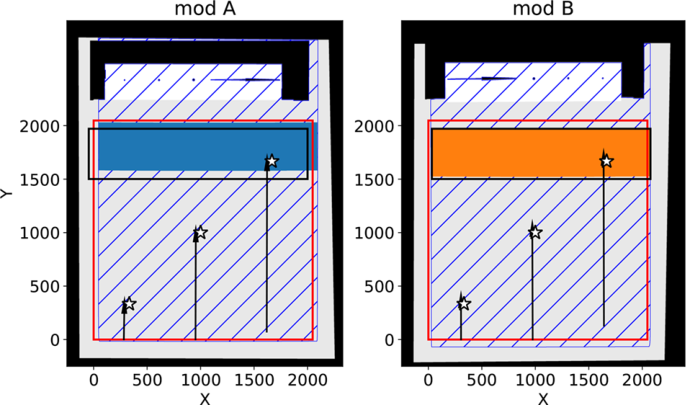}
\plotone{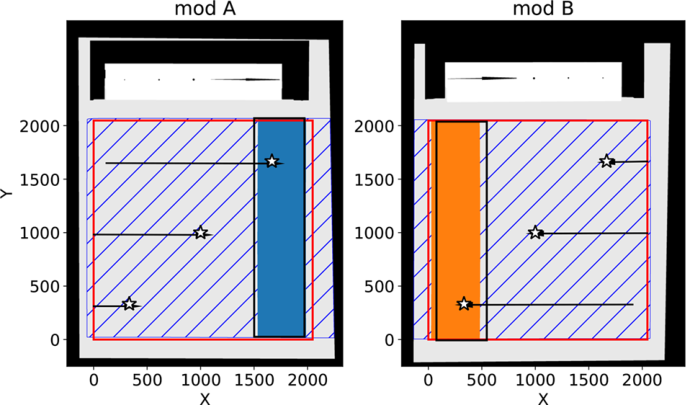}
\caption{ Location of areas for complete grism spectral coverage using F322W2. The upper row shows FOVs for the column grisms, and the lower row displays the row grisms. Diagonal lines indicate the sky area where sources can appear on the detector in grism mode. The red square outlines the detector area. Black rectangles outline
the original regions for complete spectral coverage while blue and orange indicate the updated regions for complete spectral coverage. Star symbols illustrate the fraction of the spectral range that is captured for the indicated star locations.}
\label{fig:GRISM}
\end{figure}

This NIRCam mode uses grisms in NIRCam's long wavelength arm and is complementary to the NIRISS wide field slitless mode that covers 0.8 to 2.25 $\mu m$ (\citealt{Dixon_etal2015}). The NIRCam grisms have R$\sim1200$, higher than NIRISS grisms that have R$\sim 150$. NIRCam WFSS works from 2.44 to 4.98 $\mu m$.  The short wavelength arms can be used simultaneously with the grisms to acquire a short wavelength direct image of the field. The zero deflection wavelength for NIRCam's grisms is 3.94 $\mu m$, and sources
not in the direct imaging field of view can appear in the grism field of view. This is described in more detail
in \cite{Greene_etal2017}. Commissioning data updated the spatial locations 
for best grism coverage as shown in Figure \ref{fig:GRISM}. Each module has a pair of grisms with one grism
aligned with detector rows and one aligned with detector columns.  Taking exposures with each grism can
enable better separation of potentially confused sources.

\begin{figure}[ht!]
\plotone{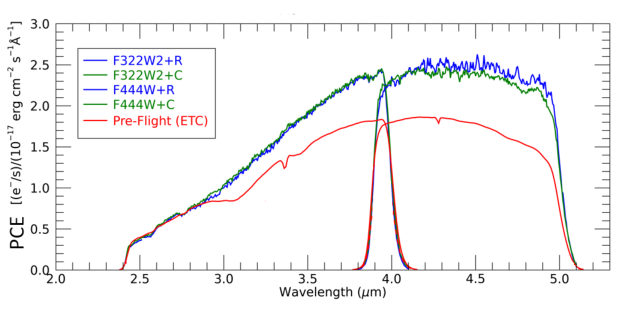}
\caption{Module A grism photon conversion efficiency derived from observations of P330-E.}
\label{fig:GRISM_ModA}
\end{figure}

\begin{figure}[ht!]
\plotone{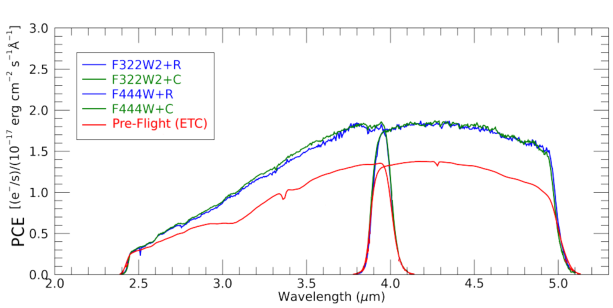}
\caption{Module B grism photon conversion efficiency derived from observations of P330-E.}
\label{fig:GRISM_ModB}
\end{figure}

\begin{figure}[ht!] 
\centering
\includegraphics[width=9cm]{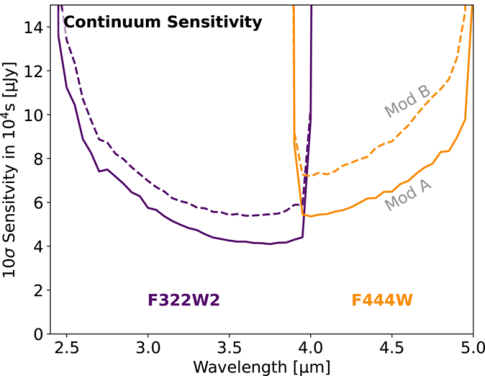}\hfill 
\includegraphics[width=9cm]{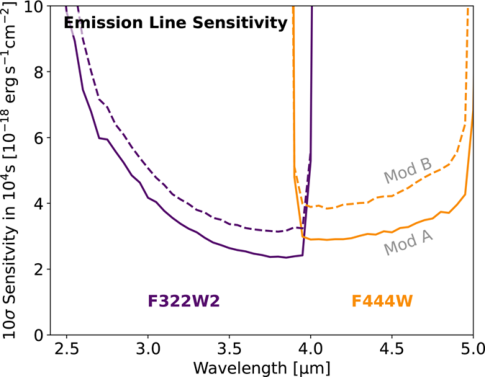}
\caption{Module A grism continuum  and line sensitivities for a $10\sigma$ detection in 10,000 seconds.}
\label{fig:cont}
\end{figure}

Because the WFSS mode mixes background photons from across the wavelength range with source photons that are
dispersed by the grism, this mode is not as sensitive as NIRSpec's slit modes. The NIRCam WFSS mode is $\sim8$ times
less sensitive than NIRSpec, but has the advantage of not requiring pre-knowledge of source positions. The utility of this mode for observing moderate to high redshift galaxies
is illustrated in \cite{Sun_etal2022a} and \cite{Sun_etal2022b} where WFSS flux calibration data taken during commissioning
revealed several $z\sim 6$ galaxies with bright emission lines and little continuum.
Figures \ref{fig:GRISM_ModA} and \ref{fig:GRISM_ModB} show that the grism 
photon conversion efficiency is higher than pre-launch predictions because there is little or no water or
other organic ices on JWST's optical surfaces. The grisms were used to check for ice contamination during 
commissioning by observing A stars with no ice absorption at $3.1 \mu m$ detected. Module A has higher throughput than Module B because the Module A grisms are anti-reflection coated
on both sides of the grisms while Module B grisms are coated only on the flat
side. Only one set of grisms were anti-reflection coated on the groove-side as there was no prior experience
demonstrating that coating grooves would not lead to problems.  Figures \ref{fig:cont} presents the $10-\sigma$ detection limits in 10,000 seconds.

\begin{figure}[ht!]
  \centering
  \includegraphics[width=9cm]{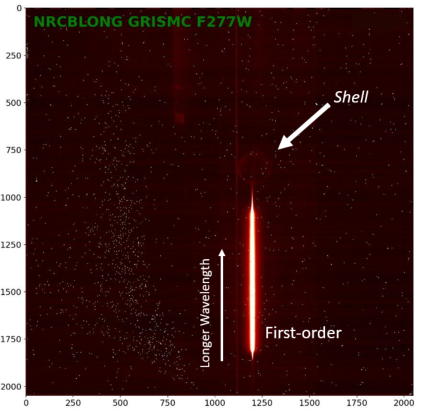}\hfill 
  \includegraphics[width=9cm]{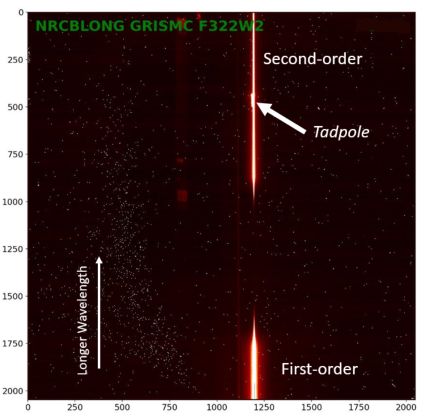}
  \caption{On the left, the \enquote{shell} grism ghost. The mechanism for this ghost was never definitively determined but likely involves multiple reflections between the grism, filter, and camera lense(s). On the right, a \enquote{tadpole} ghost is seen and is likely the 0th order grism image reflected off the filter, back to the grism, and then appearing as a 0th order image on the detector. \enquote{Tadpoles} appears at a fixed location relative to an object's 0-order image.  }
  \label{fig:g_ghosts}
\end{figure}

During testing of NIRCam before launch two types of ghost images were identified in grism data. Figure \ref{fig:g_ghosts} shows examples of these ghosts. The “tadpole” ghost appears in both Module A and Module B although it is much stronger in Module B ($>100\times$) as a result of the grisms only being anti-reflection coated on one side. The \enquote{tadpole} ghost is mostly seen with Module B column grism and appears at a fixed location relative to the 0-order image. The morphology of tadpole ghost weakly depends on the source position and filter being used. The  \enquote{shell} ghost has only been seen with Module B so far when very bright sources are observed. Either type of ghost is usually not a concern in module A unless very bright sources are observed.

\subsection{Coronagraphy Mode} \label{subsec:coronagraphy}

\begin{figure}[htbp!]
  \centering
   \includegraphics{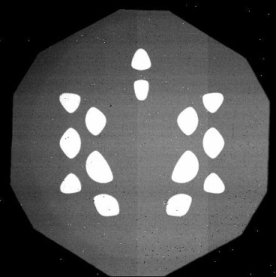}
  \caption{ An image of the Lyot stop used with round focal plane masks taken using the pupil imaging lens.}
  \label{fig:lyot}
\end{figure}

\begin{figure}[htbp!]
  \centering
   \includegraphics{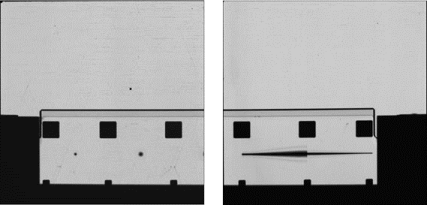}
  \caption{ View of Module A's masks through the short wavelength channel. The squares along the edge of the
  field are neutral density spots used for target acquisition.}
  \label{fig:masks} 
\end{figure}

NIRCam's coronagraphy is based on a standard Lyot-type coronagraph with five focal plane masks available. The coronagraphic focal plane masks are not in the imaging field of view unless the pupil wheel is
rotated to a Lyot stop with a wedge that deflects the field of view to include the masks. Figure \ref{fig:lyot} shows the stop used with round masks. The holes are undersized to minimize diffracted
light from the edges of the mirror segments. The throughput of the Lyot stops is $\sim 20 \% $ so throughput has been traded for better contrast. This 
optical train is shown schematically in \cite{Girard_etal2022}. Figure \ref{fig:masks}
shows the masks in a flight image against sky background. There are three round masks sized to match $6\lambda / D$ when combined with F210M ($2.1 \mu m$), F335M ($3.35 \mu m$), or F430M ($4.3 \mu m$). The choice of mask
essentially defines the inner working angle. The two bar masks provide more flexibility in terms of choice of wavelength with the mask widths varying linearly with widths corresponding to $4 \lambda / D$ from 1.7 to 2.2 $\mu m$  or 2.5 to 5 $\mu m$. The transmission of the mask 
substrate drops below 1.7 $\mu m$ so coronagraphy is not recommended at the shortest wavelengths.  The masks are not hard-edged but rather Gaussian-tapered.  Fabrication details are given in \cite{Krist_etal2009}. The masks were 
fabricated at JPL’s Microdevices Lab. \cite{Girard_etal2022} present results of coronagraphy tests performed during commissioning with performance 
exceeding the expectations given in \cite{Perrin_etal2018} with a firm detection at 3.35 $\mu m$ of the white dwarf companion HD 114174 B at a separation of 0.5 arc seconds and at a contrast of 10 magnitudes ($10^4$ fainter than the K$\sim$5.3 host star). One of the first coronagraphic observations after the start of science operations achieved
contrast of better than $10^{-5} (5\sigma) $ at 0.5 arc sec separation using the long wavelength bar mask at 
wavelengths from 2.5 to 4.6 $\mu m$ and as low as $2 \times 10^{-6}$ at 3 $\mu m$ and 1 arc sec (\citealt{Greenbaum_etal2022}).
There were no performance requirements defined for the design of the NIRCam coronagraph, but as mentioned above, 
in-flight performance exceeds pre-flight simulations in \cite{Perrin_etal2018}. As more experience is gained in the use of
the coronagraphic mode, performance improvements are likely.

\subsection{Time Series Mode}  \label{subsec:tso}

The NIRCam long wavelength grisms enable transit spectroscopy. In principle, JWST can point at a single target for up to 10 days with only brief breaks to 
reposition the downlink antenna. This capability is very different from transit work performed using the Hubble Space Telescope which has to interrupt observations every few tens of minutes unless the target is in the continuous
viewing zone. Commissioning transit observations have shown that JWST is a very good platform for observing
exoplanet transits (\citealt{Schlawin_etal2022}).  See \cite{Beichman_etal2014} for an overview of all of JWST's transit capabilities.

One module of NIRCam can be used to study objects where high time resolution data is needed.  This 
mode can use any of NIRCam's filters with one choice in the short wavelength arm and another choice in the long wavelength arm. The time resolution depends on whether full frame or subarray data are requested. In either readout choice, there is overhead from resetting the pixels used so in a typical observation the reset - read - read pattern means in one third of the time when resetting, no data are collected (this inefficiency is present also for grism
time series). The fastest sampling uses a 64 pixel x 64 pixel subarray which can be read out every 50 milliseconds.

\subsubsection{Grism Time Series}

NIRCam's grism time series capability was successfully demonstrated during commissioning. The current implementation of this mode uses a weak lens in the short wavelength arm and a grism with R$\sim 1500$ bandlimited with either the F277W, F322W2, F356W, or the F444W filter in the long wavelength arm. The grism time series mode only uses Module A. The sensitivity curve in Figure \ref{fig:GRISM_ModA} is relevant also for this mode
and shows that the F322W2 spectral range is $2.4 \mu m$ to $4 \mu m$ and is $3.9 \mu m$ to $5.0 \mu m$ for F444W. This mode can take advantage of \enquote{stripe} mode where the SCA is read out using all four output amplifiers present in the SCA but only enough rows to capture the grism spectrum. This scheme uses the row grism only. The shortest read out time is provided by using a 64x2048. This enables observing significantly brighter stars than the usual subarray mode, which reads out using only one of the SCA output amplifiers.  Table \ref{table:grism_sat}
gives the Vega magnitudes at K(2.2 $\mu m$) for several wavelengths in the spectrum using either F322W2 or F444W. The table assumes that the data are taken using a reset followed by clocking through pixels to read the pixel charge followed by a second read (RAPID mode with two groups). Using a reset followed by a single read could make the saturation limit 0.75 magnitudes brighter at the expense of the readout noise being increased by the kTC noise of$\sim 35 $ electrons.  

\begin{table}[ht]
\center{}
\caption{Vega K Magnitude for Stars Reaching 80\% of Full Well Using Stripe Mode and Reset-Read-Read}
\begin{tblr}{|c|c|c|}
\hline
Filter & $ \lambda (\mu m $)  & $K_{mag}$ \\
\hline 
 F322W2 & 2.40 &  3.99 \\
\hline
 F322W2 & 2.75 &  4.37 \\
\hline
 F322W2 & 3.23 &  4.10 \\
\hline
 F322W2 & 4.00 &  3.90 \\
\hline
 F444W & 3.90 &  3.96   \\
\hline
 F444W & 4.41 & 3.33 \\
\hline
 F444W & 5.00 & 1.99 \\
\hline
\end{tblr}
\label{table:grism_sat}
\end{table}

Several important lessons from commissioning indicate that this mode may be more user friendly than the modes used for transit observations on the Hubble Space Telescope. First, the time for detectors to settle after the start of time series observation is nearly instantaneous with a time scale of less than 4 minutes to stabilize to a level of less than 150 ppm variation ( \citealt{Schlawin_etal2022}) rather than the few tens of minutes required by WFC3IR detectors. Second, the data analysis is
more straightforward than for Hubble transit data.  The noise performance for a region of less than 30 pixels is at the expected level for photon-limited statistics. The broadband light curve averages down as $1/\sqrt{N}$ to $\sim 20$ ppm after correction for a downward trend in the signal level. Section \ref{sec:future} describes an enhancement to this mode
where the short wavelength arm could use the 0-degree DHS to acquire spectra in the 1 to 2$\mu m$ region
simultaneously with the grism spectrum.

\subsection{In Flight Anomalies} \label{subsec:anomalies}

\subsubsection{Cosmic-Rays}
\begin{figure}[htp!]
  \centering
   \includegraphics{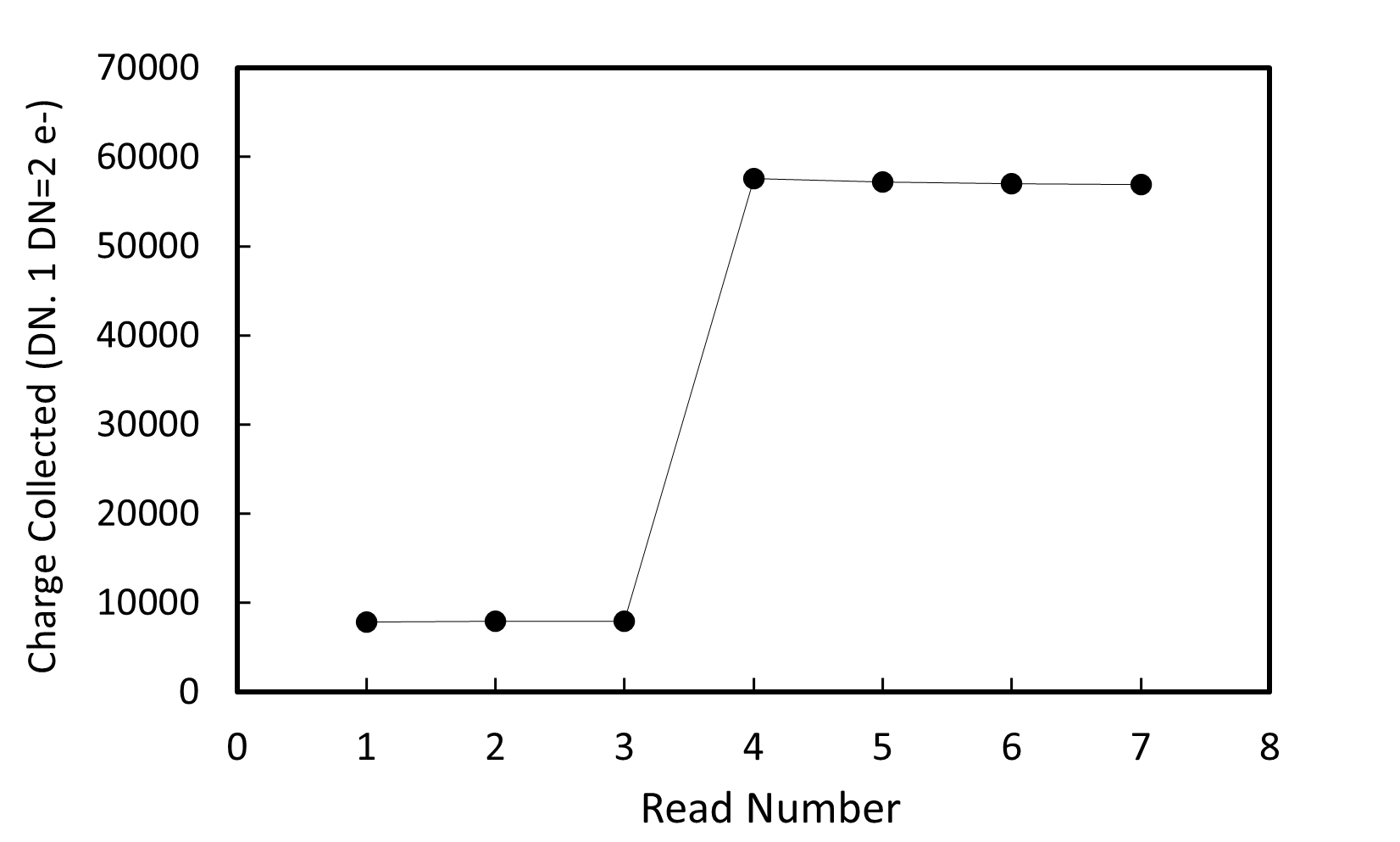}
  \caption{ Ramp from a pixel hit by a cosmic-ray.}
  \label{fig:cosmic-ray}
\end{figure}

Cosmic-rays are an expected annoyance for space instrumentation.  Figure \ref{fig:cosmic-ray} shows a ramp from a pixel with a large cosmic-ray hit that saturated the pixel between reads three and four.  The last reads show a slight decrease in charge as the charge migrates to adjacent pixels. Most hits like the one shown here affect
only one pixel and rarely a short streak of pixels. The pipeline takes these hits into account and only uses good data to derive a signal. The pipeline inserts a flag in the data quality plane in the FITS file for an image to indicate that a cosmic-ray hit was detected in a given pixel. The rate of hits matches the pre-launch estimates.

\begin{figure}[htp]
  \centering
  \includegraphics[width=9cm]{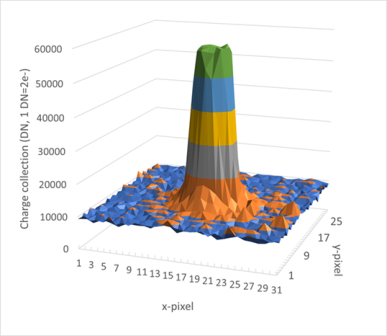}\hfill 
  \includegraphics[width=9cm]{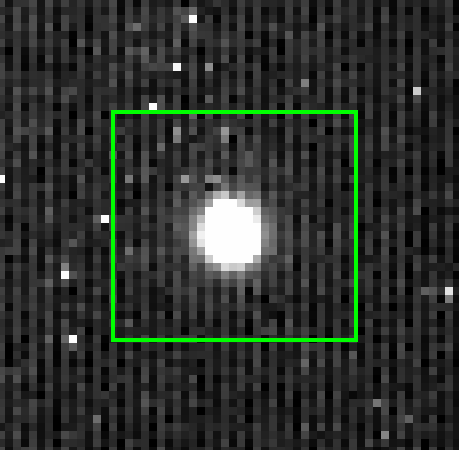}
  \caption{The right-hand image show a snowball as seen in a single frame in a ramp. The green box outlines the
  region shown in the surface plot at left.}
  \label{fig:snowballs}
\end{figure}

What caught some by surprise was the detection of \enquote{snowballs} as shown in Figure \ref{fig:snowballs}. The snowballs have appreciable spatial extent with some affecting as many as 100 pixels. Typically many of these
pixels receive a saturating level of charge as illustrated in the surface plot in Figure \ref{fig:snowballs}. Similar events had been seen in earlier space missions using HgCdTe detectors, but because the rate is quite low, they were seen rarely. Webb has a much larger number of 2Kx2K arrays so these events are more noticeable. Approximately 170 events per hour per array are seen in NIRCam arrays with roughly similar rates observed in NIRSpec and NIRISS.   There is no definitive explanation for these events. HgCdTe arrays are comprised of higher atomic weight materials than CCDs, and this suggests that a cosmic-ray could hit a heavy element nucleus and create energetic high atomic number nuclei by spallation, which then deposit significant charge in a set of pixels (G. Rieke, private communication).  

\subsubsection{Persistence}
\begin{figure}[htp!]
  \centering
   \includegraphics{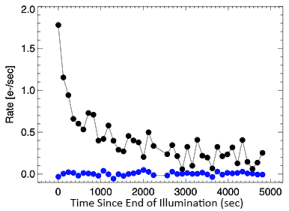}
  \caption{Persistence seen in NRCA3 . The blue points are from a dark exposure while the black points are the brightest pixel in a dark taken immediately after an on-sky image.}
  \label{fig:persist}
\end{figure}

Three of NIRCam’s SCAs, NRCA3, NRCB3, and NRCB4, exhibit noticeable persistence even several 1000s of seconds after illumination as shown in Figure \ref{fig:persist}.   A typical time between the end of one program and the start of the next is $\sim$1800 seconds so scheduling very bright source observations immediately prior to programs with long, low-signal observations should be avoided.  Saturating cosmic-ray hits, and snowballs in particular, can result in persistence images of the hit.

\subsubsection{Scattered Light and Glints}
\begin{figure}[htp!]
  \centering
  \includegraphics{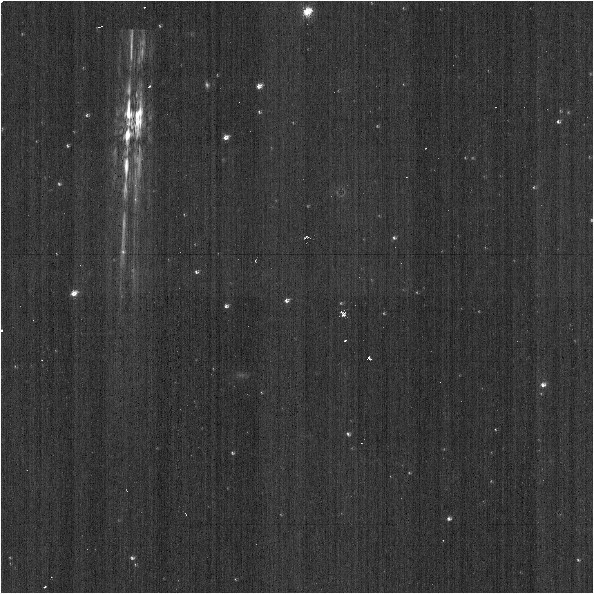}
  \caption{An example of a glint caused by star light scattering off a sharp edge in the focal plane mask.}
  \label{fig:glint}
\end{figure}

\begin{figure}[htp!]
  \centering`
  \includegraphics{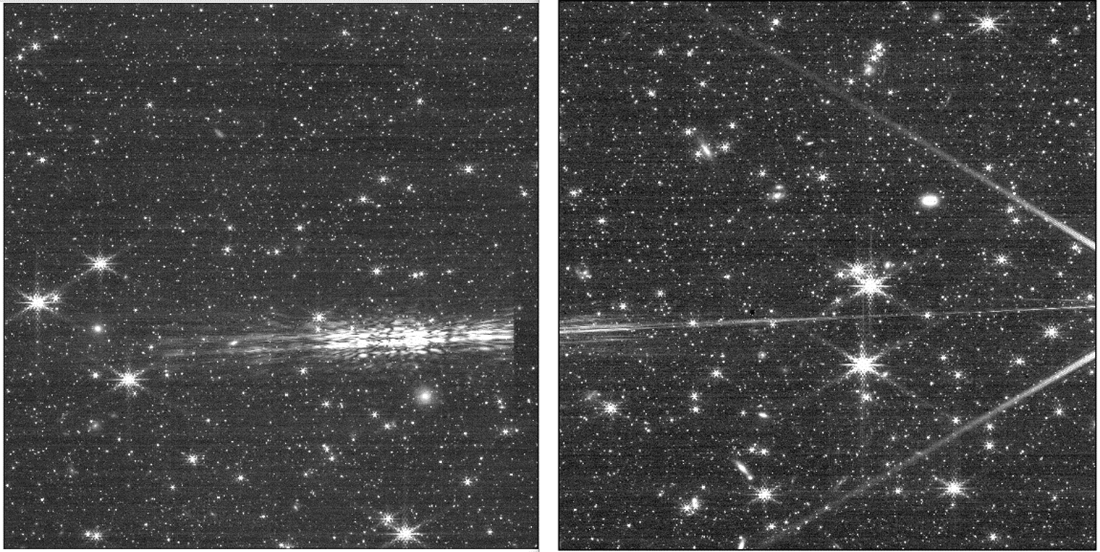}
  \caption{An example of a glint caused by star light scattering off a sharp edge inside the focal plane housing. Adjacent
  short wavelength SCAs are shown with the glint from a single star extending across both.}
  \label{fig:int_glint}
\end{figure}

Occasional glints were expected from ground tests, but some glints and some scattered light features were a surprise.
The scattered light features are described in \cite{Rigby_etal2022}, and are caused by sneak paths through
the aft optics baffle and by scattering off reflective material on the upper secondary mirror support strut.
During ground tests glints (sometimes called “Dragon’s Breath”) were discovered.  Similar artefacts have been observed in HST instruments, and result from a relatively bright point source scattering and diffracting off sharp edges.  A commissioning test collected images with a star moved across the edges of the focal plane masks on the focal plane assemblies. Figure \ref{fig:glint} shows a typical glint coming off a sharp edge in the focal plane mask, and is similar to those seen in ground test. Another type of glint is shown in Figure \ref{fig:int_glint} where light glances off a sharp edge inside the NIRCam focal plane housing.

\section{Future Improvements} \label{sec:future}
Several enhancements to the modes described in Section \ref{sec:modes} will improve observing efficiencies. These include using the DHS in parallel with the long wavelength grism for time series observations and improving coronagraphy to permit simultaneous short and long wavelength coronagraphy. Combining DHS observations with grism observations yields simultaneous wavelength coverage from $1.0\mu m$ to $2.0\mu m$ and either $2.4 \mu m$ to $4.0\mu m$  or $3.9\mu m$ to $5\mu m$ depending on the choice of long wavelength filter (\citealt{Schlawin_etal2017}, see DHS + grism data dample in this paper). This mode would not provide the continuous $1\mu m$ to $5\mu m$ coverage afforded by NIRSpec but would provide some short wavelength spectral coverage on bright ($ J < 11 $) sources which saturate in NIRSpec's prism mode. The NIRISS Single Object Slitess Spectroscopy  mode would provide more complete coverage for $\lambda < 2.5 \mu m$ but does not provide any longer wavelength coverage. \cite{Schlawin_etal2017} describe the benefits for molecular abundance retrievals using this NIRCam mode.

Acquisition of simultaneous short wavelength and long wavelength coronagraphy would use a long wavelength coronagraphic mask such as MASK335R. At 0.5 arc second separation, this mask would provide good contrast similar to the  $\sim 10^{-5}$ already achieved in the long wavelength arm and better then the contrast of $5 \times 10^{-5}$ at achieved at $2.1 \mu m$ using MASK210R (\citealt{Girard_etal2022}). These contrasts would come at the expense of loss of some inner working angle at short wavelengths compared to using MASK210R. However, for some observations such as studying debris disks, the loss of some area in the short wavelength image is far outweighed by the time savings of observing two wavelengths at once.

Simultaneous short and long wavelength coronagraphy will be available to Cycle 2 proposers, and DHS transit use will be available in the future.

NIRCam and JWST have been demonstrated to work very well, and have enabled exciting observations as revealed
by several Early Release Observations and Early Science Observations.



\begin{acknowledgments}
Many people at Lockheed Martin's Advanced Technology Center contributed to the NIRCam Project. We especially want to thank Alison Nordt, Eric Dixon, Tony Magoncelli and Liz Osborne for support 
during commissioning. The development, testing, and commissioning of NIRCam was funded by NASA Contract NAS5-02105. A portion  of this research was  carried out at the Jet Propulsion Laboratory, California Institute of Technology, under a contract with the National Aeronautics and Space Administration (80NM0018D0004). D.J.\ is supported by NRC Canada and by an NSERC Discovery Grant. We also want to acknowledge the contributions of three NIRCam team members who passed away during the development  of NIRCam. Chad Engelbracht analyzed much of the detector test data that were used
to select the flight parts. John Stauffer devised the
original calibration plan for NIRCam. Don Hall provided advice on many issues and detector production in particular.
The referee is thanked for helpful suggestions.

\end{acknowledgments}


{}



\end{document}